# Symmetry-Broken Kondo Screening and Zero-Energy Mode in the Kagome Superconductor $CsV_3Sb_5$

Yubing Tu[1,2,8], Zongyuan Zhang[1,2,3,8], Wenjian Lu[4,8], Tao Han[1,2], Run Lv[4], Zhuying Wang[5], Zekun Zhou[1], Xinyuan Hou[1,2,3], Ning Hao[6✉], Zhenyu Wang[5,7✉], Xianhui Chen[5,7], Lei Shan[1,2,3,7✉]

[1]Information Materials and Intelligent Sensing Laboratory of Anhui Province, Institutes of Physical Science and Information Technology, Anhui University, Hefei, 230601, China.

[2]Leibniz International Joint Research Center of Materials Sciences of Anhui Province, Anhui University, Hefei, 230601, China.

[3]Center of Free Electron Laser and High Magnetic Field, Anhui University, Hefei, 230601, China.

[4]Key Laboratory of Materials Physics, Institute of Solid State Physics, HFIPS, Chinese Academy of Sciences, Hefei, 230031, China.

[5]Department of Physics, CAS Key Laboratory of Strongly-coupled Quantum Matter Physics, University of Science and Technology of China, Hefei, Anhui 230026, China

[6]Anhui Provincial Key Laboratory of Low-Energy Quantum Materials and Devices, High Magnetic Field Laboratory, HFIPS, Chinese Academy of Sciences, Hefei, 230031, China.

[7]Hefei National Laboratory, University of Science and Technology of China, Hefei 230088, China

[8]These authors contributed equally: Yubing Tu, Zongyuan Zhang, Wenjian Lu.

✉Email: haon@hmfl.ac.cn; zywang2@ustc.edu.cn; lshan@ahu.edu.cn

The quantum states of matter reorganize themselves in response to defects, giving rise to emergent local excitations that imprint unique characteristics of the host states. While magnetic impurities are known to generate Kondo screening in a Fermi liquid and Yu-Shiba-Rusinov (YSR) states in a conventional superconductor, it remains unclear whether they can evoke distinct phenomena in the kagome superconductor $AV_3Sb_5$ (where A is K, Rb or Cs), which may host an orbital-antiferromagnetic charge density wave (CDW) state and an unconventional superconducting state driven by the convergence of topology, geometric frustration and electron correlations. In this work, we visualize the local density of states induced near various types of impurities in both the CDW and superconducting phases of $CsV_{3-x}M_xSb_5$ (M = Ta, Cr) using scanning tunneling microscopy. We observe Kondo resonance states near magnetic Cr dopants. Notably, unlike in any known metal or CDW compound, the spatial pattern of Kondo screening breaks all in-plane mirror symmetries of the kagome lattice, suggesting an electronic chirality due to putative orbital loop currents. While Cooper pairs show relative insensitivity to nonmagnetic impurities, native V vacancies with weak magnetic moments induce a pronounced zero-bias conductance peak (ZBCP). This ZBCP coexists with trivial YSR states within the superconducting gap and does not split in energy with increasing tunneling transmission, tending instead to saturate. This behavior is reminiscent of signature of Majorana zero modes, which could be trapped by a sign-change boundary in the superconducting order parameter near a V vacancy, consistent with a surface topological superconducting state. Our findings provide a new approach to exploring novel quantum states on kagome lattices.

Impurities in matter are not always undesired entities. Single-atom impurities offer a significant window into the quantum state of the host material through associated local excitations. A well-known example is the detrimental effect of nonmagnetic impurities on Cooper pairs in unconventional superconductors [1-3]. In the case of magnetic impurities, their coupling with either the itinerant electrons or Cooper pairs results in Kondo singlets or Yu-Shiba-Rusinov (YSR) states [4], respectively, leaving behind a characteristic spatial structure predominantly influenced by the lattice potential [5-9]. In principle, the local electronic structure around impurities should also be shaped by the symmetry of the host electronic orders, thereby providing a fundamental fingerprint of the symmetry-broken quantum state. However, to date, no substantial evidence of order-induced symmetry-broken Kondo or YSR states has been observed, although their intensity may vary slightly in real space [10, 11]. Furthermore, magnetic impurities play a vital role in generating new quantum functionalities: They can induce global ferromagnetic order in topological insulators, enabling the realization of the quantum anomalous Hall effect [12, 13], and create Majorana zero modes (MZMs) on the surface of connate iron-based topological superconductors [14, 15]. Therefore, it is of fundamental importance to explore the single-atom impurity state in quantum materials with novel electronic orders and band topology.

Recently, the V-based kagome family $AV_3Sb_5$ has garnered tremendous attention due to its rich electronic phases [16-19]. These materials undergoes a charge density wave (CDW) transition at $T_{CDW}$ ~ 100 K, with a 2×2 lattice reconstruction in the kagome plane, followed by the emergence of superconductivity at $T_c$ ~ 1–3 K. Since the vanadium d-orbitals form a series of van Hove singularities at energies close to the Fermi level [20], long-range Coulomb interactions promoted by sublattice interference are predicated to generate various exotic states [21-24]. Indeed, a set of experiments has revealed additional electronic crossover within the CDW phase, accompanied by electronic nematicity [25-28], unidirectional $4a_0$ order [29] or time-reversal symmetry breaking [26, 30-35]. Of particular interest is the time-reversal symmetry breaking, as it is associated with the long-sought loop-current phase [36, 37], rather than any spin order. However, the signal for this order is extremely small, leading to conflicting experimental results [38-40]. Similarly, the nature of the superconducting state remains controversial: while pair density wave [41, 42] and time-reversal symmetry breaking [43] suggest exotic pairing mechanisms, photoemission spectroscopy and irradiation-based studies imply phonon-mediated conventional superconductivity [44, 45]. In addition, the $Z_2$ band topology may contribute to nontrivial excitations in the superconducting state [16, 46, 47]. Therefore, exploring the response of these electronic states to single-atomic impurities is crucial for gaining microscopic insight into their nature, an endeavor that remains to be pursued.

We experimentally close this gap by investigating the local excitations near various impurities in both the CDW and superconducting states of $CsV_{3-x}M_xSb_5$ (M= Ta, Cr) by utilizing an ultra-low temperature STM. Studies of impurity effects in crystals typically suffer from either weak scattering or unidentified impurities. In this work, we introduce impurity concentrations in a controlled manner, enabling us to quantify their number from STM images and, coupled with first-principles calculations, determine their nature. Different from previous studies, we focus on V-site dopants in the kagome plane (Fig. 1a,b), which is expected to induce a stronger scattering potential for the host states. Our key observations include the presence of a fully symmetry-broken Kondo screening near magnetic Cr dopants as well as the coexistence of robust topologically-protected zero-energy bound states (ZBSs) and YSR states centered at weak-magnetic V vacancies. The former indicates a strong interaction between Kondo screening and the CDW order at low temperatures, possibly involving chirality induced by the chiral flux phase [48-52]. The latter suggests that the magnetic moment near a V vacancy can induce a sign change in the superconducting order parameter, potentially trapping MZMs associated with $Z_2$ topological surface states.

## Results and discussions

The crystalline structure of $CsV_3Sb_5$ is depicted in Fig. 1a,b, with vanadium atoms forming a two-dimensional kagome network. Antimony atoms occupy two different lattice sites: the hexagonal centers of the kagome network (Sb2) and the honeycomb lattice above and below the kagome layer (Sb1). Figure 1c,e shows typical topographies of cleaved $CsV_{3-x}M_xSb_5$ (M = Ta, Cr) samples, with a bias voltage of $V_s$ = +1.5 V, displaying Sb1 termination. The bright protrusions on the Sb-Sb bridges indicate the locations of dopants in the underlying kagome layer, similar to previous reports [28]. The dopant concentrations, x, in $CsV_{3-x}M_xSb_5$ are determined to be 0.075 and 0.003 for Ta and Cr, respectively (Supplementary Fig. 1), which agrees well with the results from energy dispersive x-ray spectroscopy and inductively coupled plasma atomic emission spectroscopy measurements (Supplementary Table 1).

These diluted substitutions slightly suppress the CDW transition temperature. For comparison, the resistivity data are displayed in Supplementary Fig. 2, showing that $T_{CDW}$ ($T_c$) for CsV$_{2.925}$Ta$_{0.075}$Sb$_5$ and CsV$_{2.997}$Cr$_{0.003}$Sb$_5$ are determined to be 84 K (3.0 K) and 91 K (2.6 K), respectively. Figure 1d,f depicts STM topographies taken in the same field of view as those in Fig. 1c,e, respectively, but with much lower biases. The Fourier transformations are shown in the insets of Fig. 1d,f, confirming that both the 2×2 CDW order and the unidirectional 4×1 order on the Sb surfaces are preserved in these Ta-doped and Cr-doped samples.

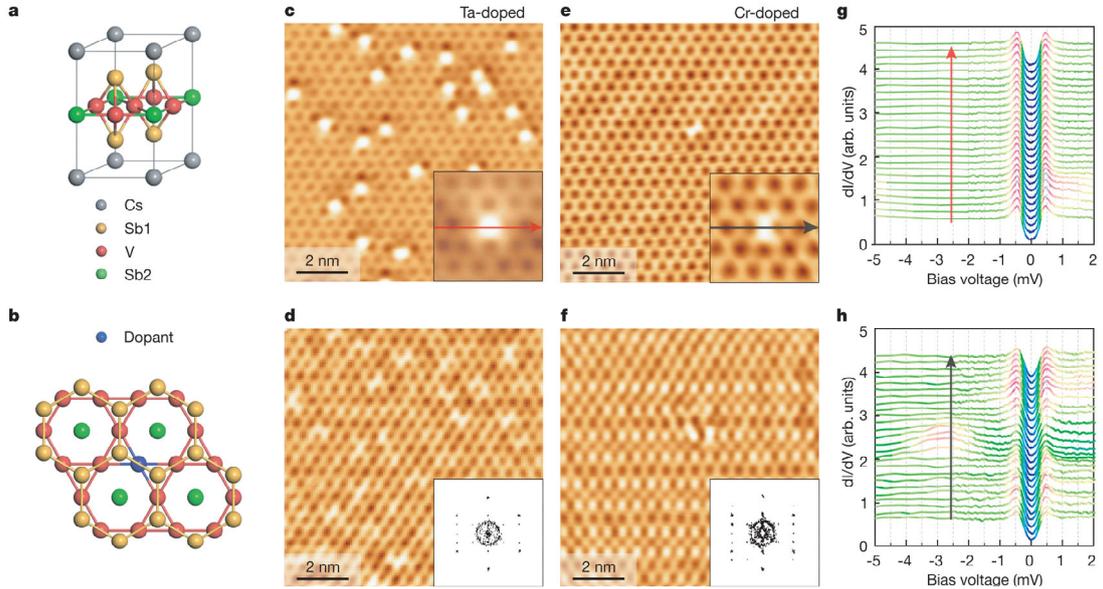

**Figure 1. Identification of Ta, and Cr dopants. a,** Side view of the crystal structure of doped CsV$_3$Sb$_5$ viewed from the side. **b,** Top view of the crystal structure, showing the Ta or Cr atoms (blue) substituting the V atoms (red) in the kagome layer. **c-f,** Typical topographies on the Sb-terminated surface of Ta-doped CsV$_3$Sb$_5$ (**c, d**) and Cr-doped CsV$_3$Sb$_5$ (**e, f**) taken at positive high bias voltages (**c, e**) and negative low bias voltages (**d, f**). The insets in (**c**) and (**e**) show the Sb1 honeycomb lattice with a dopant located halfway between a pair of Sb1 atoms. The insets in (**d**) and (**f**) show the Fourier transforms of the topographies, revealing charge modulations. **g, h,** The d$I$/d$V$ linecuts along the arrows in (**c**) and (**e**). The STM setup conditions were sample bias $V_s$ = +1.5 V, tunneling current $I_t$ = 400 pA (**c**); $V_s$ = +1.4 V, $I_t$ = 50 pA (**c**, inset); $V_s$ = -65 mV, $I_t$= 20 pA (**d**); $V_s$ = +1.5 V, $I_t$ = 100 pA (**e**); $V_s$ = +1.5 V, $I_t$ = 100 pA (**e**, inset); $V_s$ = -80 mV, $I_t$ = 100 pA (**f**); $V_s$ = -5 mV, $I_t$ = 300 pA, bias modulation $V_m$ = 0.1 mV, temperature $T$ = 0.4 K (**g**); $V_s$ = -5 mV, $I_t$ = 300 pA, $V_m$ = 0.1 mV, $T$ = 0.4 K (**h**).

We first investigate the impact of non-magnetic Ta impurities on the superconducting state. As shown in Fig. 1g, tunneling spectra of d$I$/d$V$ acquired along a linecut across a single Ta dopant exhibit an almost identical spectral shape. This suggests that a non-magnetic impurity does not hinder superconductivity or induce any visible in-gap states. In addition to the intentionally introduced Ta impurities, we observed two types of inherent Sb vacancies and found that they are not pair-breakers (Supplementary Fig. 3). The immunity of Cooper pairs to non-magnetic impurities on both V and Sb sites supports the scenario of s-wave pairing, as a sign-change gap structure would be more sensitive to defect scattering according to conventional theories. However, a recent theoretical study points out that the particular sublattice character of the kagome network renders disorder only a weak pair-breaker for any singlet pairing, regardless of sign changes in the gap structure [53]. Thus, strictly speaking, no definitive conclusion can be drawn based on these nonmagnetic impurities at the current stage, and future work is warranted. Consequently, our subsequent investigations will focus on magnetic impurities.

Chromium dopants are generally regarded as magnetic impurities in conventional superconductors. In contrast to the spectra around non-magnetic impurities, a conductance peak is notable outside the superconducting gap in the spectra taken along a linecut across a Cr dopant, as shown in Fig. 1h. Analysis of more data obtained from dozens of Cr dopants reveals that this conductance peak has a characteristic energy varying between -2.5 meV and +5.5 meV (supplementary Fig. 4), which is very close to the Fermi energy, as expected in the picture of Kondo resonance surrounding a magnetic impurity embedded in a bath of conducting electrons. Figure 2a displays the morphology of another Cr dopant and its corresponding spectrum, which also exhibits a conductance peak near the Fermi level, indicated by the arrow in Fig. 2b. This resonance peak can be well fitted with the Fano function [54], and the Kondo resonance width $\Gamma$ follows a well-known temperature dependence, yielding a Kondo temperature of $T_K$ ~18 K. To further confirm the Kondo origin of this peak, we measured its evolution under an out-of-plane magnetic field. As shown in Fig. 2c, the peak broadens at lower fields and eventually splits at high magnetic fields, fully consistent with Zeeman splitting and a Landé g-factor of approximately $g$ ~ 2.2 for a Kondo resonance state [55, 56]. With the local moment being screened, we find no visible YSR bound states inside the superconducting gap near the Cr dopant in our study. In contrast, YSR states have been observed in surface-deposited Cr clusters, while no Kondo resonance has been reported [57]. These differences suggest that, although $k_B T_K$ (~ 1.55 meV) is significantly larger than $\Delta$ (~ 0.3 meV) for Cr atoms doped into the V lattice (supplementary Fig. 5), the Cr clusters deposited on the surface Sb layer have a much lower Kondo energy and therefore experience weaker Kondo screening.

After determining the Kondo resonance nature of the peak in d$I$/d$V$ near the Cr impurities, we next examine its characteristic spatial structure. A real-space conductance map obtained at the resonance energy is shown in Fig. 2d, with spectra taken along the arrowed line shown in Fig. 2e. Surprisingly, the spatial pattern of the Kondo state exhibits a spatial oscillation that extends over several nanometers (the largest extension known to date) along one of the three lattice directions and is not centered at the impurity site, which differs from the Kondo pattern observed in any known metals or CDW compounds. Specifically, this pattern clearly breaks the vertical and diagonal mirror symmetries of the kagome plane. The lowered symmetry becomes more apparent when we plot the distribution of intensity maxima relative to the lattice. Figure 2f illustrates four equivalent legs surrounding an impurity site (the red dot) by merely considering the geometry of the kagome lattice. These legs correspond to a two-fold symmetric environment of a single V atom. However, Kondo features could only be observed experimentally on one leg, as indicated by the light blue dots. Moreover, although the features are centered at the triangles of the kagome lattice, the strongest intensity is located at the triangle next-nearest to the impurity (refer to Fig. 2d).

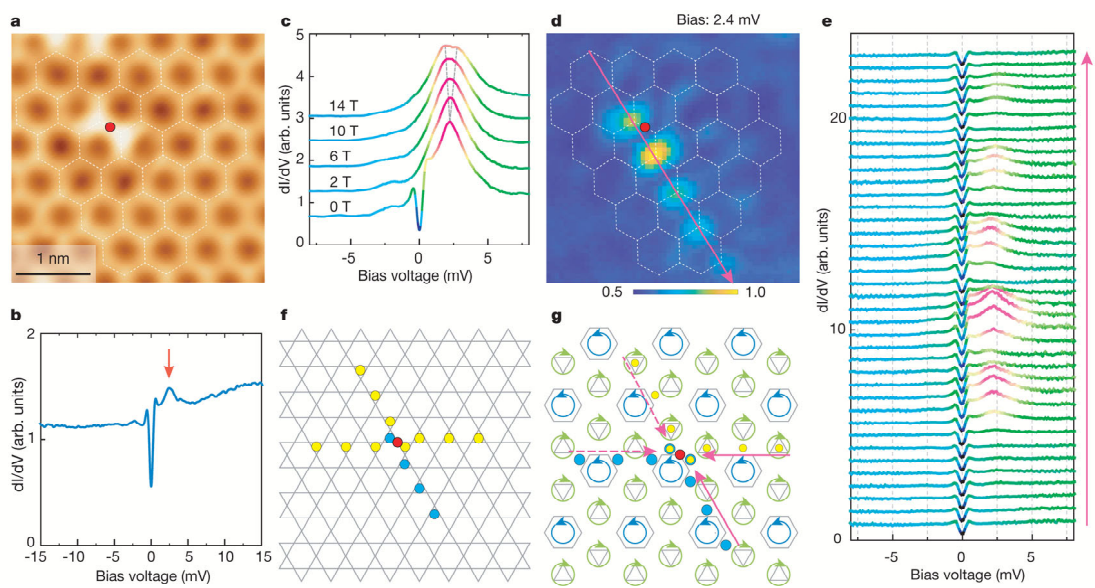

**Figure 2. Kondo state observed around a Cr impurity. a,** STM topography of a single Cr dopant marked by a red dot. **b,** Spectrum taken on the top of the Cr dopant. **c,** Spectra taken near the Cr dopant under different magnetic fields. **d,** Zero-energy conductance map obtained in the same region

as shown in (**a**). **e**, Spectra taken along the arrowed line in (**d**). **f**, Schematic of the expected spatial extension of the Kondo state based on lattice symmetry, manifesting as an X-shaped pattern comprised of four legs. The single leg observed experimentally is denoted by the light blue dots. **g**, The same lattice as that in (**f**), but taking into account the chiral flux phase with a 2×2 CDW order. The arrows on the circles indicate the directions of current flux loops. The STM setup conditions were $V_s$ = -65 mV, $I_t$ = 20 pA (**a**); $V_s$ = -20 mV, $I_t$ = 300 pA, $V_m$ = 0.2 mV, $T$ = 0.4 K (**b**); $V_s$ = -10 mV, $I_t$ = 300 pA, $V_m$ = 0.15 mV, $T$ = 0.4 K (**c**); $V_s$ = -15 mV, $I_t$ = 300 pA, $V_m$ = 0.15 mV, $T$ = 0.4 K (**e**).

The Kondo screening of a single magnetic impurity, whether an atom or molecule, depends on the spatial distribution of the magnetic moment [58, 59] and its exchange coupling to the itinerant electrons within a specific crystal lattice [5, 6, 60]. Consequently, the real-space pattern of the Kondo resonance should be determined by both the magnetic moment and the crystal lattice symmetries. In our case, the extension of the Kondo resonance with its maximum intensity away from the impurity suggests that the local orbital of Cr atoms is unlikely to be the simple reason for the spatial asymmetry (see Fig. 4f for the calculated distribution of the local moment of Cr atoms). Therefore, it is necessary to carefully check the electronic orders present here. Figure 2g depicts the tri-hexagonal configuration of the V-kagome net in the CDW state, overlapped by the same legs as shown in Fig. 2f. From a symmetry standpoint, the two upper legs (smaller yellow dots) and the two lower legs (larger light blue dots) are not equivalent. However, further symmetry breaking into a single leg suggests the involvement of an additional electronic order. By measuring the spatial feature of approximately twenty Cr atoms, we can rule out the surface 4×1 reconstruction and the nematic CDW as possible origins for the single leg, since the orientation of the Kondo state pattern frequently switches among the three lattice directions in one $C_2$ domain (Supplementary Fig. 6).These patterns are neither related to long-range strain in the lattice within the field of view (Supplementary Fig. 7). In this context, the lack of in-plane mirror symmetry with respect to the location of the defect gives rise to chirality in the kagome plane, either from the structure or the electronic degree of freedom. Given that the low-temperature structural distortion is very weak [25, 61, 62], it seems unlikely that subtle deviations in bond length are the cause of the dramatic asymmetry in Kondo screening. Alternatively, by considering the opposite directions of the current flux loops experienced by the injecting quasiparticles towards the impurity along different legs or paths, a further symmetry breaking can be seen between the two pairs of legs indicated by the dashed lines and solid lines, respectively, as illustrated in Fig.

2g. Therefore, the most likely candidate contributing to the fully symmetry-broken Kondo state observed here is a chiral CDW order or chiral flux phase, which aligns with the observation of electronic magnetochiral anisotropy in transport studies [34].

While the above discussions provide a plausible explanation for the unprecedented symmetry-broken Kondo state, we note that a microscopic model of the interplay between the local magnetic moment and the chiral flux phase, i.e., orbital antiferromagnetism, is still needed. There are also open questions that require clarification: first, the strong intensity oscillation occurs within the lattice constant period; second, the highest intensity is found in the next-nearest neighbor triangle, rather than the nearest neighbor triangle, relative to the impurity site in the kagome lattice (Fig. 4f); and third, the Kondo state displays a consistent symmetry-broken pattern regardless of variations in its resonance energy (Supplementary Information Section 4). These unusual features of the Kondo state observed in the kagome superconductor $CsV_3Sb_5$ thus present an opportunity to explore new aspects of Kondo physics involving intricate interactions between Kondo screening and other electronic correlations.

Next, we move to the native V vacancies, which are evident in both pristine and doped samples. These vacancies manifest as dumbbell shapes in the topographic images under negative scanning biases, as shown in Fig. 3a. Conversely, they transform into dark bars when a positive bias is applied, as depicted in Fig. 3b. The bias-dependent alterations in topography can be well replicated by density functional theory (DFT) computations (Supplementary Fig. 8), validating their identification as V vacancies. Notably, unlike non-magnetic defects such as Ta and Sb, a conspicuous zero-bias conductance peak (ZBCP) emerges in spectra proximate to V vacancies (Fig. 3c), accompanied by a suppression of the superconducting coherence peaks—an indication of the presence of ZBSs.

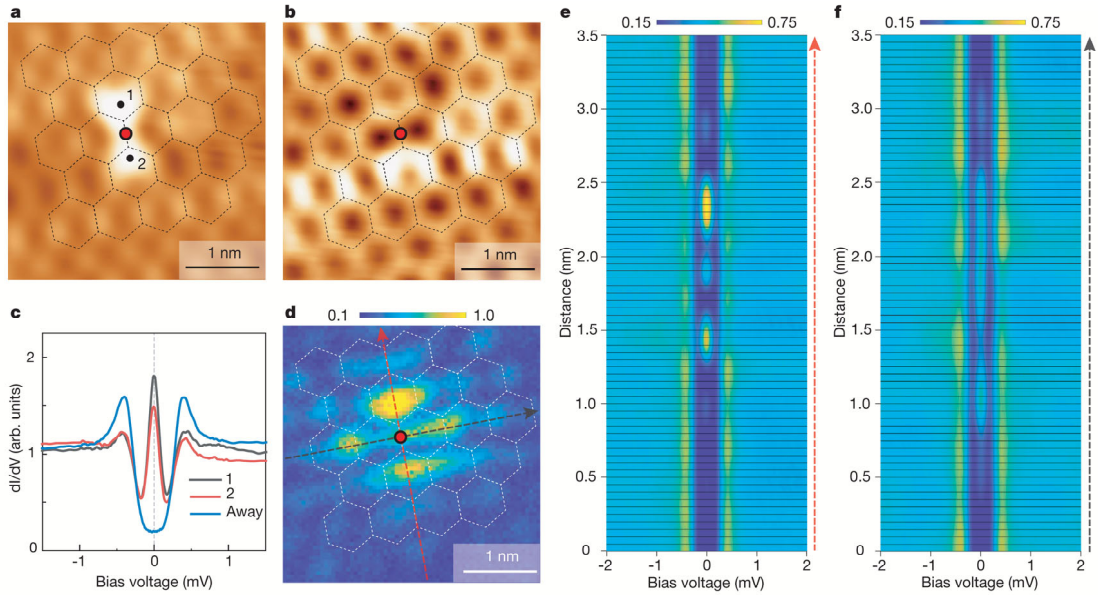

**Figure 3. Spatial evolution of the impurity state induced by a V vacancy. a, b,** STM topographies of a single V vacancy taken at negative (**a**) and positive (**b**) low bias voltages. The center of the V vacancy is marked by a red dot. **c**, Spectra taken on and away from the top of dumbbell-shaped V vacancy. **d**, Zero-energy conductance map for the V vacancy as shown in (**a**). **e, f,** Spectra taken along the lines in (**d**). The STM setup conditions were $V_s$ = -80mV, $I_t$= 20 pA (**a**); $V_s$ = +80mV, $I_t$= 200 pA(**b**); $V_s$ = -5 mV, $I_t$ = 300 pA, $V_m$= 0.05 mV, $T$ = 0.4 K (**c**); $V_s$ = -5 mV, $I_t$ = 300 pA, $V_m$ = 0.1 mV, $T$ = 0.4 K (**e**); $V_s$ = -5 mV, $I_t$ = 300 pA, $V_m$ = 0.1 mV, $T$ = 0.4 K (**f**).

To elucidate the unique attributes of these ZBSs, we conducted temperature and magnetic field-dependent spectral analyses on V vacancies (Supplementary Fig. 9). Interestingly, the ZBCP fades above 1.2 K, a temperature well below the critical temperature ($T_c \sim$ 3 K), while it diminishes progressively with escalating magnetic field until reaching the upper critical field, devoid of any discernible energy shift or splitting, ultimately coinciding with the disappearance of the superconducting coherence peaks. Spectroscopic imaging around V vacancies in a pristine sample (Fig. 3d) uncovered that the ZBCP predominantly exhibits a four-lobed cloverleaf pattern in real space, extending up to 2nm from the defect site. This observed two-fold symmetry aligns with the local environment of the V sites in the kagome lattice. For enhanced clarity, tunneling spectra traversing the defect center along the primary symmetry axes are presented in Fig. 3e,f, revealing an oscillatory variation of the ZBCP along both axes without significant energy splitting.

Subsequently, we compared the dI/dV spectra in close proximity to V vacancies with those further afield. Figure 4a-c summarizes findings from three distinct V vacancies:

alongside the ZBSs, a pair of particle-hole symmetric features, manifesting as kinks or shoulders (highlighted by arrows), is conspicuous in Fig. 4a,b but absent in Fig. 4c. DFT calculations further revealed that V vacancies induce relatively weak magnetic moments on neighboring V atoms (Fig. 4d,e; see Supplementary Information Section 7 for comprehensive details), hence attributing the energy-symmetric kinks to the YSR states. The characteristic energy and spectral weight of these YSR states vary across different defects. Nevertheless, the persistence of ZBCP suggests that despite coexistence, the ZBS and YSR states may originate differently (supplementary Fig. 10).

In order to get further insight into the zero-energy states, we have investigated their evolution upon increasing junction transmission (i. e., the tip-to-sample distance). It has been proposed that the binding energy of YSR states undergoes a systematic shift due to enhanced exchange coupling as the tip nears the sample [63-66], whereas MZMs remain pinned at zero energy [67] and are expected to exhibit a quantized conductance of $2e^2/h$ for an individual MZM under high junction transmission [68-70]. The outcomes in proximity to V vacancies are illustrated in Figs. 4g,h. As the junction transmission $G_N$ (defined as $I_t/V_s$) increases, we found that the ZBCP remains fixed at zero energy. Notably, upon reaching a certain value of approximately $0.7 \times 2e^2/h$, the ZBCP begins to saturate towards $0.8 \times 2e^2/h$, whereas the high-bias conductance continues to escalate, surpassing the ZBCP (Fig. 4g). Our conductance measurements were taken with the tip positioned at point "1" in Fig. 3a, corresponding to the brightest lobe in Fig. 3d. Given that the ZBSs spatially divide into four lobes as seen in Fig 3d, a single lobe's contribution of $0.8 \times 2e^2/h$, close to a quantized conductance, seems improbable unless the number of ZBSs exceeds one. We will subsequently attribute this phenomenon to the presence of a pair of energy-degenerate ZBSs. The temperature, magnetic-field, and junction-transmission dependencies of these ZBSs resemble those reported for potential MZMs in several connate topological superconductors [68, 69], despite the anticipated disruption from quasiparticles owing to the small superconducting gap.

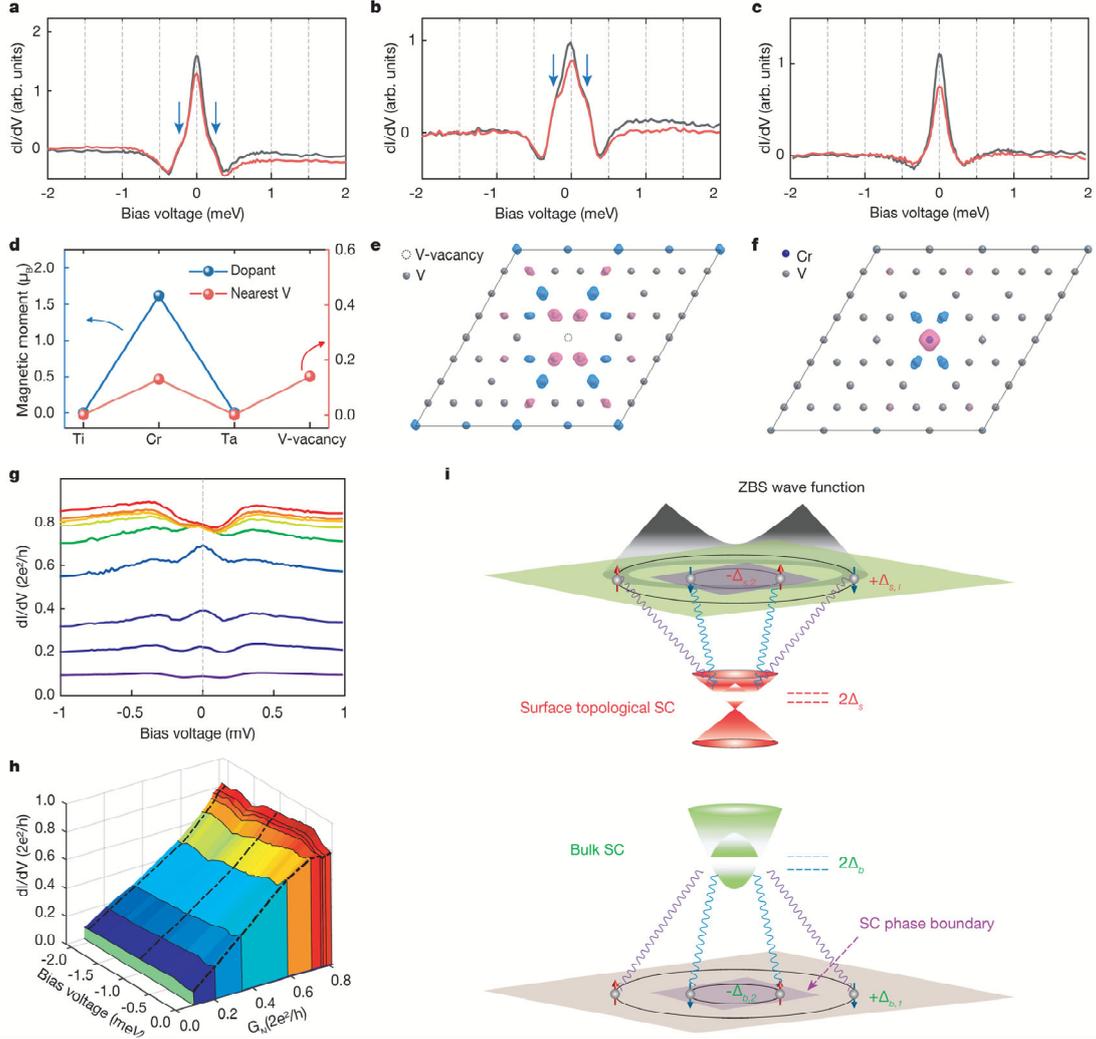

**Figure 4. Explanation of the ZBS around a V vacancy. a-c**, Spectra acquired on the two lobes of three individual dumbbell-shaped V vacancies with backgrounds subtracted to emphasize the in-gap states. Blue arrows indicate the peaks of YSR states. The backgrounds were taken on the clean Sb surface approximately 30 Å away from the vacancies. **d**, Magnetic moments calculated for different dopants and V vacancy in the kagome layer. Results for both the dopant sites and nearest-neighbor V sites are presented for comparison. **e,f**, Calculated real-space spin density isosurfaces for $CsV_3Sb_5$ with V defects and Cr doping. The isosurface values are set to 0.002 $\mu_B \cdot Å^{-3}$ and 0.006 $\mu_B \cdot Å^{-3}$, respectively. Pink and blue colors represent spin-up and spin-down components. **g**, A set of dI/dV spectra with increasing junction transmission (from the bottom to the top) via reducing the tip-sample distance. **h**. A three-dimensional plot of junction transmission-dependent $dI/dV$ spectra. **i**. Schematic diagram for the generation of topologically protected ZBSs. For the bulk superconductivity (SC, $\Delta_b$), the V atoms nearest to the vacancy with finite magnetic moments as shown in (**e**) can trigger a quantum phase transition with a phase boundary separating two sign-reversed superconducting regimes (bottom panel of **i**). Surface topological SC occurs in the topmost Sb atom layer through the superconducting proximity effect via V-Sb interlayer coupling (top panel of d). The spatial configuration of the superconducting order parameter ($\Delta_{s,1}$, $\Delta_{s,2}$) of surface topological SC mirrors that of the bulk SC ($\Delta_{b,1}$, $\Delta_{b,2}$). A pair of robust ZBSs emerges at the phase boundary in surface topological SC and the relevant wave functions are schematically illustrated (top panel of i). The STM setup conditions were $V_s$ = -5 mV, $I_t$ = 300 pA, $V_m$ = 0.05 mV, $T$ = 0.4 K

(**a-c**); $V_s$ = -2 mV, $V_{mod}$ = 0.05 mV, $T$ = 0.4 K (**g, h**).

It is noteworthy that iron impurities in materials like Fe(Te,Se) [67], LiFeAs [71] and PbTaSe$_2$ [72] can induce Majorana-like ZBSs. Theoretical scenarios have been proposed for the generation of robust ZBSs, including quantum anomalous vortices [15] and phase boundary modes in topological surface superconductors [73, 74]. Given the small magnetic moments near V vacancies, weak spin-orbit coupling, and the extremely low concentration of vacancies—coupled with the above-mentioned conductance data implying a pair of ZBSs—we advocate for the latter scenario. In the context of the kagome superconductor studied herein, the elevated density of states near van Hove singularities at the Fermi level could facilitate a quantum phase transition cooperated by the magnetic moments adjacent to a V vacancy [75]. This transition enables the superconducting order parameter to change sign around the nearest-neighbor V atoms to the vacancy, even with small magnetic moments. The physical picture can thus be reduced to the issue of bound states in surface topological superconductivity with 0-π phase boundaries [73, 74], naturally elucidating the emergence of a resilient ZBS pair [73, 74], as depicted in Fig. 4i. Furthermore, the coexistence of trivial YSR states and robust ZBSs becomes intelligible. Specifically, the magnetic nearest-neighbor V atoms to the vacancy give rise to trivial YSR states, while the superconducting phase boundary—an outcome of the quantum phase transition—manifested in the surface topological superconducting state, engenders the topologically protected ZBSs.

## Conclusion

In this study, we have utilized localized states around defects to explore novel electronic states in the kagome superconductor CsV$_3$Sb$_5$. By substituting a vanadium atom with a chromium atom in the kagome plane, we observed a clear Kondo resonance state, indicating a strong exchange coupling between the magnetic moment and the itinerant electrons. However, the Kondo pattern spreads in real space and breaks all mirror planes of the kagome lattice with respect to the defect, in which contrasts with any known metal or CDW compound. These lower-symmetry spatial structures suggest electronic chirality due to putative orbital loop currents, providing the first glimpse into how orbital antiferromagnetic order with chiral hopping of electrons couples with a magnetic moment. This observation calls for further investigation.

Furthermore, we discovered that small magnetic moments near V vacancies act detrimentally on Cooper pairs, inducing a pronounced and robust ZBCP inside the superconducting gap. The evolution of this ZBCP under varying temperature, magnetic field, and junction transmission suggests the possible presence of MZMs around the defect. We propose that the robust ZBSs can be generated at the superconducting phase boundary, manifested in the surface topological superconducting state, positioning $CsV_3Sb_5$ as a promising platform for studying MZMs. Overall, our results strongly indicate that the behavior of quasiparticle bound states from various dopants and vacancies in kagome superconductors is highly dependent on charge orders, specific electron correlations, band topology and superconductivity. These findings open new avenues for understanding and manipulating quantum states in kagome metals.

Online content

Any methods, additional references, Nature Research reporting summaries, source data, extended data, supplementary information, acknowledgements, peer review information; details of author contributions and competing interests; and statements of data and code availability are available at supplementary materials.

## Methods
### Crystal growth

$CsV_3Sb_5$-derived single crystals were synthesized via a flux method as previously reported [76]. Starting materials with a stochiometric molar ratio (Cs: M+V: Sb) of 1:3:15 were mixed in an alumina crucible and then sealed in a quartz tube. The mixture was slowly heated to 1273 K and maintained at this temperature for 24 hours. The quartz tube was then slowly cooled to 1073 K at a low cooling rate of 1-2 K/h before being centrifuged at a speed of 3000 r/min. Electrical transport measurements were performed using a Quantum Design Physical Property Measurement System (PPMS DynaCool).

### STM measurements

STM/STS experiments were carried out using an ultra-low temperature scanning tunneling microscope (USM-1300, Unisoku Co., Ltd., Japan). All samples were cleaved in an ultrahigh vacuum at 77 K, and then transferred to the STM head, which was maintained at a low temperature around 2 K. Electrochemically etched tungsten tips were used, following electron-beam heating and calibration on the Au(111) surface.

Tunneling spectra were acquired using a standard lock-in technique with a bias modulation of $V_{mod} = 0.05$ mV at 973.127 Hz, unless otherwise specified.

**Density functional theory**

The first-principles calculations based on density functional theory (DFT) were performed using the Vienna Ab initio Simulation Package (VASP) [77, 78]. The exchange-correlation function was described by the generalized gradient approximation (GGA), parameterized by Perdew-Burke-Ernzerhof (PBE) [79]. The plane-wave kinetic cutoff energy was set to 400 eV. A 4×4×1 supercell was used to simulate the M atom (M = Ti, Cr, and Ta) doped and V-defected $CsV_3Sb_5$. The self-consistent calculations were performed with a 5×5×11 $k$-point grid and an energy convergence of $10^{-6}$ eV. To simulate STM topographies, we constructed a 4×4×1 superstructure with an Sb-surface using a 20 Å vacuum layer and performed DFT calculations with a 9×9×1 $k$-point grid. The STM images were calculated using the Tersoff-Hamann method [80], implemented within the VASP code.

## Data availability

The data supporting the findings of this study are available from the corresponding author upon reasonable request. Source data are provided with this paper.

**Acknowledgements** This work was supported by the National Key R&D Program of China (Grant No. 2022YFA1403200, 2024YFA1611103), the Innovation Program for Quantum Science and Technology (Grant No. 2021ZD0302802), the National Natural Science Foundation of China (Grant No. 12474128, 92265104, 12204008, 12104004, 12304162, 12374133, 12022413), National Nature Science Foundation of China under Contracts No. U2032215, the Basic Research Program of the Chinese Academy of Sciences Based on Major Scientific Infrastructures (Grant No. JZHKYPT-2021-08), Anhui Provincial Major S&T Project(s202305a12020005).

**Author contributions** L. S. designed the experiments. Y. T. and Z. Z performed STM experiments and data analysis with guidance from L. S. and Z. W. T. H. prepared and characterized the samples. W. L. and R. L. carried out theoretical calculations. L. S., N. H., X. H., Z. W. and X. C. interpreted the results and wrote the manuscript. All of the authors discussed the experimental data and commented the manuscript.

**Competing interests** The authors declare no competing interests.